\documentclass[twocolumn, a4paper]{revtex4}
\usepackage{graphicx}
\usepackage{dcolumn}
\usepackage{amsmath}
\usepackage{color}
\begin{document}
\hsize\textwidth\columnwidth\hsize\csname@twocolumnfalse\endcsname

\title{Development and operation of the twin radio frequency single electron transistor for cross-correlated charge detection}
\author{T. M. Buehler, D. J. Reilly, R.  P. Starrett, N. A. Court, A.  R.
Hamilton, A.   S. Dzurak and R. G. Clark \vspace{0.3cm}}

\affiliation{Centre for Quantum Computer Technology, Schools of
Physics and Electrical Engineering \& Telecommunications,
University of New South Wales, Sydney 2052,
Australia\vspace{0.5cm}}


\begin{abstract}

\noindent Ultra-sensitive detectors and readout devices based on
the radio frequency single electron transistor (rf-SET) combine
near quantum-limited sensitivity with fast operation. Here we
describe a twin rf-SET detector that uses two superconducting
rf-SETs to perform fast, real-time cross-correlated measurements
in order to distinguish sub-electron signals from charge noise on
microsecond time-scales. The twin rf-SET makes use of two tuned
resonance circuits to simultaneously and independently address
both rf-SETs using wavelength division multiplexing (WDM) and a
single cryogenic amplifier. We focus on the operation of the twin
rf-SET as a charge detector and evaluate the cross-talk between
the two resonance circuits. Real time suppression of charge noise
is demonstrated by cross correlating the signals from the two
rf-SETs. For the case of simultaneous operation, the rf-SETs had
charge sensitivities of $\delta q_{SET1} = 7.5 \mu e/\sqrt{Hz}$
and $\delta q_{SET2} = 4.4 \mu e/\sqrt{Hz}$.

\end{abstract}

\maketitle

\section{Introduction}

The radio frequency single electron transistor (rf-SET) is a
nano-scale device capable of detecting fractions of an electron
charge on sub-microsecond time-scales \cite{Schoelkopf_science}.
In addition to being an exquisite electrometer
\cite{schoelkopf_nature}, the rf-SET may also find application as
a single photon \cite{Komiyama,schoelkopf_submil} and particle
detector \cite{Bouchiat}, micro-mechanical displacement sensor
\cite{Blencowe}, ultra-high impedance voltage amplifier
\cite{Segall,Stevenson_amp,Stevenson_APL} and ultimately as a fast
single-shot read out device in solid-state quantum computers
\cite{Aassime_PRL}. At present the sensitivity of rf-SETs is
primarily limited by the post-amplifier, which is typically a GaAs
high-electron mobility transistor (HEMT) operating at $T \sim$ 4K.
In the near future, this limitation is likely to be overcome via
the use of rf amplifiers based on a micro-strip dc superconducting
quantum interference device (SQUID) which have demonstrated noise
temperatures near 100mK at frequencies suitable for rf-SET
operation \cite{Bradley}. At low frequencies (below 100kHz),
rf-SETs suffer from 1/f-type charge noise associated with the
motion of charge in the SET dielectric tunnel barriers and
substrate. In particular telegraph noise, that results from charge
fluctuators that are strongly coupled to the SET, poses a
challenge to the widespread use of rf-SETs as both detectors and
readout devices.

In an effort to address the issue of telegraph noise we have
utilized two superconducting rf-SETs to form a cross-correlation
charge detector, termed a {\it twin} rf-SET. The cross-correlation
technique is advantageous in suppressing charge noise originating
from fluctuating traps in the surrounding material system and
constitutes a means of discriminating read out signals from
spurious charge fluctuations on micro-second time-scales
\cite{Buehler_APL}. Independent control of the source-drain bias
across each superconducting SET permits both devices to
simultaneously operate at the point of maximum sensitivity and
measurement efficiency \cite{clerk_PRL}. By making use of
wavelength division multiplexing (WDM) the twin rf-SET provides
two (or more) outputs for cross-correlation measurements using
only a single cryogenic following amplifier and a single
transmission line. Such multiplexed arrangements are highly
desirable in instances where arrays of rf-SETs are used as
detectors \cite{Stevenson_APL}.
\begin{figure}[t!]
\begin{center}
\includegraphics[width=7.5cm]{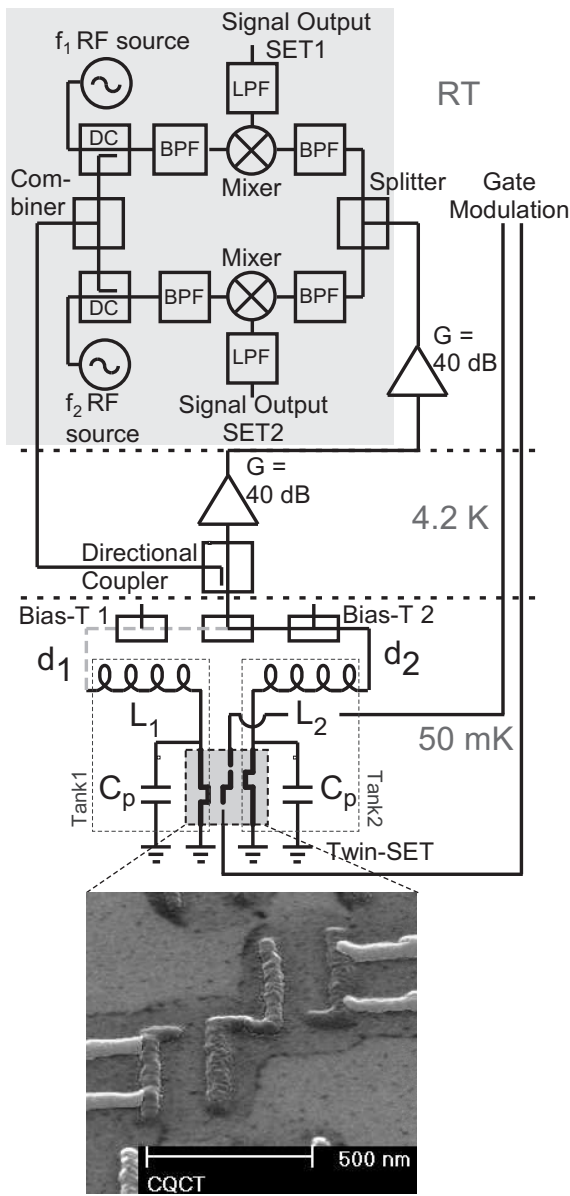}
\caption{Schematic of the twin rf-SET setup. Signals are
coupled to the tank circuits by using a directional coupler and
are then directed to the corresponding SET by the tuned impedance
transformers. The reflected signal feeds cryogenic (gain = 40dB)
and room temperature (gain = 40dB) amplifiers. Two bias-tees allow
independent dc-biasing of each SET. \textbf{Shaded region:}
Schematic of the rf carrier generation and signal demodulation.
The incident carrier wave is produced by combining the output of
two independent rf signal generators. {\bf The bottom section}
shows a SEM image of a twin-SET device. The tunnel barriers
required for Coulomb blockade are formed after the first
evaporation step by in-situ oxidation of the aluminum surface. For
SET2 (right) the overlaps of source and drain leads (second
evaporation) with the SET island (first evaporation) can be
seen}
\-vspace{-0.5cm}
\end{center}
\end{figure}

The article is organized as follows: in Sec. II we briefly review
the rf-SET. Sec. III introduces the twin rf-SET and the
multiplexing technique. We also discuss our demodulation method
that makes use of a mixer-based detection scheme. In Sec. IV we
discuss the operation of the resonant circuits and focus on the
degree of cross-talk between the circuits and the issues that
affect independent operation of the SETs. Additionally we present
electromagnetic (EM) simulation results of our circuit and show
good agreement with our data. Sec. V presents frequency and time
domain data for the case of simultaneous operation. We demonstrate
charge sensitivities of  $\delta q_{SET1} = 7.5 \mu e/\sqrt{Hz}$
and $\delta q_{SET2} = 4.4 \mu e/\sqrt{Hz}$ and show how the
technique of cross-correlation can be used to suppress spurious
charge noise on microsecond time-scales.

\section{rf-SET}

Normal state SET charge detectors are based on the Coulomb
blockade \cite{natobook} of electrons tunneling across two low
capacitance junctions $J_{1,2}$  ($C_{1},R_{1}$ and
$C_{2},R_{2}$). An `island' with a characteristic charging energy
$E_{C} = e^{2}/2C_{\Sigma}$, where $C_{\Sigma}$ is the capacitive
coupling to the environment, is formed between the tunnel
junctions. The sequential electron tunneling current across the
island can be controlled by the voltage on a gate which is
capacitively coupled to the island. Coulomb blockade is effective
for temperatures $T < E_{C}/k_{B}$ and for tunneling resistances
$R_{1,2}$ of order of the resistance quantum $h/e^{2} \sim$
26k$\Omega$. These restrictions generally limit the operation of
SETs to cryogenic temperatures. Traditionally, the bandwidth of
SETs has been limited to a few kHz by the large $RC$ time constant
associated with the SET output resistance and the capacitance of
the wiring from the SET to the room temperature amplifiers. In
contrast to the conventional SET, the rf-SET
\cite{Schoelkopf_science} makes use of a $LC$ impedance
transformer to match the impedance of the SET to the
characteristic impedance of a transmission line. In this regime,
either the reflected \cite{Schoelkopf_science} or transmitted
\cite{NTT} power of an incident rf carrier wave is a function of
the SET resistance. Mapping the device resistance to changes in
the reflected or transmitted power thereby allows fast charge
detection. In the case of reflection measurements, the incident rf
carrier is coupled to the $LCR$ matching network via a directional
coupler and the reflected power is coupled to a low noise
cryogenic HEMT amplifier. The $LCR$ circuit is described by the
loaded quality factor $Q = (Z_{LC}/R_{SET} + Z_0/Z_{LC})^{-1}$,
where $Z_0$ is the characteristic impedance of the transmission
line and $R_{SET}$ is the differential resistance of the SET and
$Z_{LC} = \sqrt{L/C}$. The (loaded) $Q$ factor is inversely
proportional to the bandwidth of the impedance matching network.

SETs fabricated from metals that exhibit superconductivity at
these temperatures generally also display rich current - voltage
characteristics in association  with the interplay between
charging effects and Cooper pair transport across junctions
$J_{1}$ and $J_{2}$. Notably, in the vicinity of the threshold
voltage for current transport (defined by the superconducting gap
and the charging energy), resonant tunneling processes can be
observed which are governed by the dynamics of coherent Cooper
pair and incoherent quasi-particle transport \cite{Fulton_89}.
Superconducting SETs biased to a Josephson quasi-particle
resonance (JQP) have recently been explored in conjunction with
reading-out a quantum computer and shown to approach quantum
limits for measurement efficiency \cite{clerk_PRL}.

\section{Twin rf-SET setup}

The technique of wavelength division multiplexing has long been
applied in communications engineering as a means of utilizing
channel resources efficiently. In line with earlier work on SQUIDs
\cite{squids_WDM}, recent work by Stevenson {\it et al.,}
\cite{Stevenson_APL} has explored the possibility of using
multiplexed rf-SETs to amplify signals from high impedance photon
detectors for application in astronomy. Here we extend these
results and note that for our case where the SETs are
independently biased using bias-tees, a lumped-element circuit
analysis is generally inadequate in describing our data.
Consequently we present circuit simulation results that account
for the length of transmission lines used in our setup
\cite{microwave_office}.

Turning now to the specifics of our setup, we achieve multiplexing
of signals from the two rf-SETs via two tuned $LC$ impedance
transformers (Figure 1). Each tank circuit can be approximated by
a parasitic capacitance $C$ to ground and a chip inductor with
inductance $L_{1,2}$. The circuits are tuned by choosing
appropriate inductances to transform the impedance of the SET
(typically 40 - 100k$\Omega$) downward towards a characteristic
impedance of 50$\Omega$ at the respective resonant frequency
$\omega_{1,2} = 1/\sqrt{CL_{1,2}}$. Figure 1 shows a schematic of
our cryogenic setup, including two separate bias-tees for
independent dc-biasing of each SET. In this setup the length of
coaxial transmission line between the `T' section and the SETs
(see Figure 1, $d_{1}$ dashed line and $d_{2}$ solid line) is a
critical parameter determining the cross-talk between the SET tank
circuits. The `T' arrangement can be considered a single shunt
`stub', where the length of the stub and transmission line between
the load and the stub position tune the admittance seen looking
into the line \cite{Pozar}. In the present experimental
arrangement we place our cryogenic HEMT amplifier at the $T$ = 4K
stage of our dilution refrigerator to avoid the use of cryogenic
feed-throughs and to reduce losses between the SET and the
amplifier. The twin rf-SET data presented here was taken using
tank circuits with resonance frequencies $f_{1} \sim 335$MHz and
$f_{2} \sim 360$MHz, using inductances of 780nH and 660nH, with a
total parasitic capacitance estimated to be $\sim 0.3$pF for each
SET. The choice of frequencies corresponds to the bandwidth where
our cryogenic rf-amplifier \cite{Berkshire} maintains optimum
noise performance and in addition, where chip inductors are
readily available. Variations in the parasitic capacitance $C$
(and resonance frequency $\omega$) for different samples can be
attributed to changes in the length of bond-wires from the
inductors to the sample contact pads, changing $C$ by tens of
$fF$.
\begin{figure}[t!]
\begin{center}
\includegraphics[width=8.5cm]{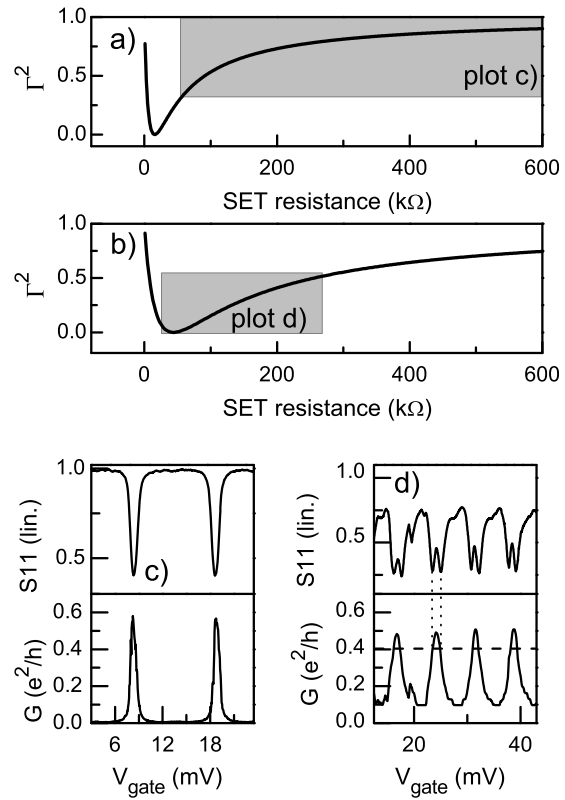}
\caption{\textbf{a) and b)} Lumped element calculations
of the power reflection coefficient as a function of the SET
resistance for impedance transformers designed to match
17k$\Omega$ to $Z_0 = 50\Omega$ for a) and  50k$\Omega$ to $Z_0$
for b). The shaded regions correspond to the matching regimes   
shown in c) and d). \textbf{c)} Reflected rf power (top, $S11$) 
and dc conductance (bottom) in the superconducting state. The SET
resistance remains above 17k$\Omega$ (note the different scale on
the y-axis) throughout and the impedance transformer operates in 
the under-matched regime. \textbf{d)} Reflected rf power (top,   
$S11$) and dc conductance (bottom) with the SET biased to a region
where the resistance at the top of a peak is less than
50k$\Omega$. The non-monotonic dependence on gate bias arises as
the operating point moves from under to over matched.}
\vspace{-0.5cm}
\end{center}
\end{figure}

The incident carrier wave is produced by combining the output of
two independent rf signal generators (Figure 1). Each generator
output is fed to a directional coupler in order to tap off power
(-16dB) for the SETs. The remaining signal (-0.11dB) is directed
to the mixers as a reference signal. A subsequent
splitter/combiner is used to feed the two frequencies to one
semi-rigid coaxial waveguide into the cryostat. After the signal
enters the cryostat the rf power is coupled to the tank circuits
using another directional coupler. The signals $f_{1,2}$ are
directed to SET$_{1,2}$ by the tuned impedance transformers,
tank$_{1,2}$. The reflected signal is then coupled to the
cryogenic amplifier (40dB gain) and a room temperature amplifier
(40dB gain). A power splitter feeds the reflected rf signal to two
mixers for demodulation. High roll-off, narrow bandpass filters
\cite{reactel} are used before both the local oscillator (LO) and
RF mixer inputs to suppress inter-modulation distortion
\cite{Pozar} and higher order components. Active phase-shifters
are used to ensure constructive interference of RF and LO inputs
at the mixer. Finally, the intermediate frequency (IF) mixer
product is low-pass filtered before being output into either a
multichannel oscilloscope or spectrum analyzer for time and
frequency domain measurements respectively. To further improve
signal quality we also use various attenuators and filters at room
and cryogenic temperatures.
\begin{figure}[t!]
\begin{center}
\includegraphics[width=8.5cm]{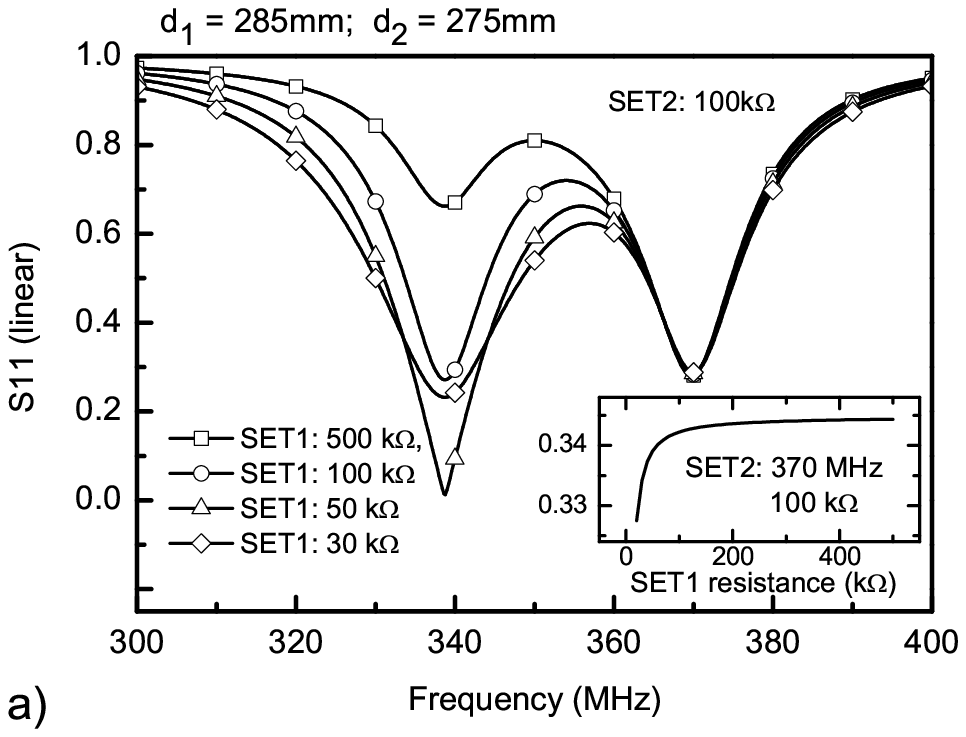}\\
\includegraphics[width=8.5cm]{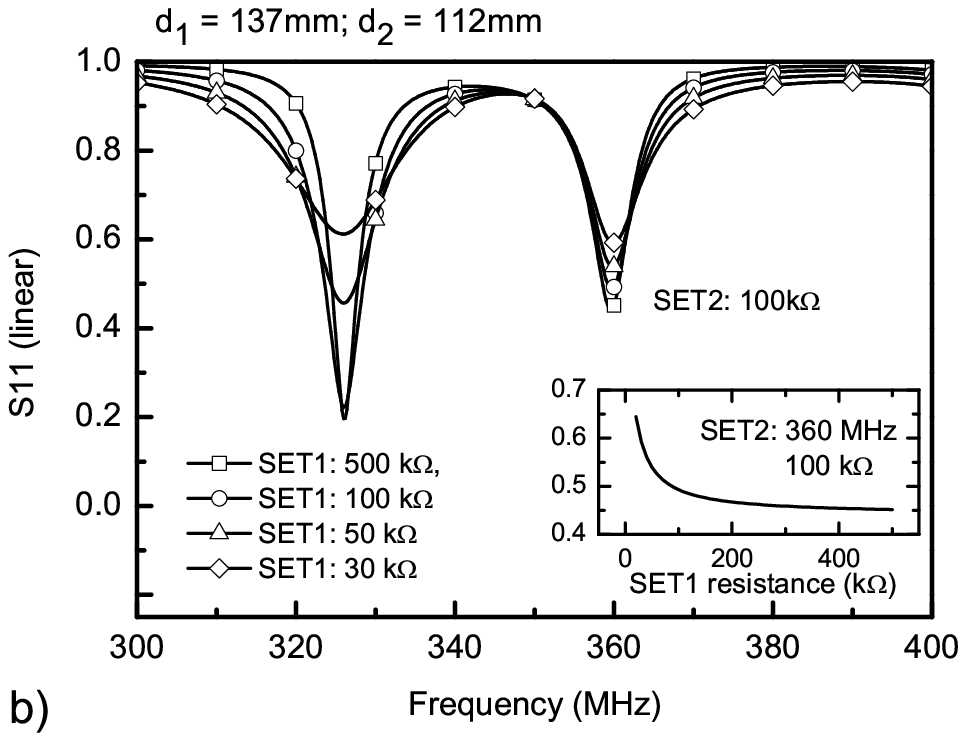}  
\caption{Simulation results \cite{microwave_office}
showing the effect of shunt stub tuning. \textbf{a)} Shows the
ideal case where cross-talk is kept to a minimum ($d_1$ = 285mm
and $d_2$ = 275mm). As indicated by the inset, $S11$ at $\omega_2$
is hardly effected by varying the resistance of SET1, only varying
slightly in the over-matched case. Contrasting this behavior
\textbf{b)} shows simulation results for an unoptimized stub
network. In this scenario the cross talk between the two tank
circuit is near 50\% despite the increased $Q$ factor. For clarity
the {\it linear} S11 is shown.}
\vspace{-0.5cm}

\end{center}
\end{figure}

The lower part of Figure 1 shows a SEM image of an
$Al/Al_{2}O_{3}$ twin-SET device used in these experiments. The
device was fabricated using electron beam lithography and the
standard shadow mask evaporation technique \cite{dolan}. The
tunnel junctions are formed where the leads (horizontal) overlap
with the `islands' (vertical). In this arrangement the SETs are
$\sim $500nm apart and control gates used in these experiments
couple to both SETs inducing charges of similar magnitude on both
devices. In addition there is a central double-dot structure,
developed for read out simulations but not relevant to the
measurements described here \cite{Buehler_APL}.

\section{Tank circuit characterization}

In order to maximize the sensitivity of the rf-SET, the device
should be operated in a regime where the amount of reflected power
depends most strongly on the SET resistance. Figures 2a and 2b
show the calculated power reflection coefficient $\Gamma^{2}$
($\Gamma = (Z_{0} - Z_{LCR})/(Z_{0} + Z_{LCR})$) in a
\textit{single} rf-SET arrangement as a function of the SET
resistance for two different matching networks. For Figure 2a the
tank circuit $LC$ values were chosen to transform a 17k$\Omega$
impedance to $Z_0$ (50$\Omega$) at resonance and for Figure 2b to
transform 50k$\Omega$ to $Z_0$. For perfect matching the power
reflection coefficient then approaches zero when the SET
resistance is 17k$\Omega$ for figure 2a and 50k$\Omega$ for figure
2b. We discuss two modes of operation, the under-matched (or
under-coupled) regime, where the device resistance is greater than
the matching resistance and the over-matched (or over-coupled)
regime where the device resistance is smaller than the matching
resistance. Over-matching yields an enhanced depth of modulation
in association with the large slope of $\Gamma ^{2}$ as a function
of the SET resistance, but reduces the $Q$ factor \cite{Roschier}.

Figures 2c and 2d show measurement results for a single rf-SET.
The reflected rf power (top) is measured together with the dc
conductance (bottom) as a function of gate voltage. The data shown
in Figure 2c is limited to the under-matched regime, where the
resistance of the SET is always greater than 17k$\Omega$. The
corresponding behavior of the reflection coefficient is indicated
in Figure 2a by the darker shaded region. In this regime the
reflected power increases monotonically with increasing SET
resistance, so that the $S11$ data shown in Figure 2c is a mirror
image of the SET conductance. In contrast, the data shown in
Figure 2d was taken with the SET biased to a point where the
differential resistance is less than  50k$\Omega$ at the top of a
Coulomb blockade peak. Moving from the blockaded state to a point
where the differential resistance is less than 50k$\Omega$ results
in a non-monotonic response in the reflected rf power as shown in
the upper section of Figure 2d. This behavior is due to the
impedance transformer operating initially in the under-matched
regime and moving to over-matching as the differential resistance
moves through 50k$\Omega$, as indicated by the dark shaded region
in Figure 2b. The non-monotonic behavior of the transfer function
in the over-matched regime complicates operation of the device as
a charge detector, as seen in the upper panel of Figure 2d.
Over-matching also reduces the $Q$-factor so that in a multiplexed
arrangement the resonances need to be separated further in
frequency to minimize cross-talk.

Turning now to simultaneous operation of both SETs we note that if
there was no cross-coupling between the two circuits, then the
output signal for tank2 should be independent of the resistance of
SET1. Clearly cross-talk is a function of the overlap between the
two resonance traces defined by their `single circuit' $Q$
factors, so that the resonance frequency of the two circuits
should be well separated to minimize cross-talk. In spite of well
separated resonances however, additional effects can strongly
modify the cross-talk between the two SETs. We now discuss the
dominant mode of cross-talk in our setup. In the multiplexed
arrangement shown in Figure 1, the branching `T' section behaves
as a stub tuning network since at the resonant frequency of SET1,
tank2 appears as a load at the end of a length of transmission
line in parallel with the transmission feed-line to SET1. The
impedance seen looking into the `T' section is \cite{Pozar},
$1/Z_{in} = 1/Z_1 + 1/Z_2$, where

\begin{equation}
Z_{1,2}(\omega_{1,2}, d_{1,2}) = Z_0 \frac{Z_{tank1,2} + j Z_0 tan
\beta d_{1,2}}{Z_0 + j Z_{tank1,2} tan \beta d_{1,2}}
\end{equation}
and $\beta = 2 \pi / \lambda_{1,2}$. For the case where the tank
circuits are well separated in frequency, with tank2 close to an
open circuit for $\omega_1$,
\begin{equation}
Z_2 = -j Z_0 cot \beta d_2
\end{equation}

so that the impedance $Z_{1,2}$ at $\omega_1$ looking into the `T'
section is just determined by $d_1$, $d_2$ and $Z_{tank1}$. In an
ideal arrangement $d_2$ should be of zero length, or adjusted so
that the impedance $Z_2$ appears as an open line. If this
condition is achieved  the amount of reflected power is solely
determined by $Z_{tank1}$. In the case where this is not achieved,
the stub geometry and length of the transmission lines between the
`T' and the SET tank circuits strongly affect both the impedance
matching of the $LC$ network to the SETs and the degree of
cross-talk between the two circuits. Figure 3a and 3b compare the
two cases of `length matching' using an EM simulation software
package \cite{microwave_office}. Optimum length matching is
illustrated in Figure 3a where $S11$ is plotted as a function of
frequency for different values of SET1 resistance, including both
under-matching (SET $>$ 50k$\Omega$) and over-matching (SET $<$
50k$\Omega$). The resistance of SET2 is held constant at
100k$\Omega$ and inductances $L_1$=780nH, $L_2$=660nH define the
resonances at $\omega_1$=339MHz and $\omega_2$=370MHz for a
parasitic capacitance $C$=0.3pF. With transmission line lengths
set at $d_1$ = 285mm and $d_2$ = 275mm, near independent operation
of the SETs can be achieved. The inset shows the variation in S11
at $\omega_2$ as a function of SET1 resistance. Only in the
over-matched case ($R_{SET1} = 30k\Omega$) does the resistance of
SET1 weakly affect the reflected power at $\omega_2$.  Turning to
the unoptimized case illustrated in Figure 3b ($d_1$ = 137mm and
$d_2$ = 112mm) we firstly note that the variation in line length
has altered the resonant frequencies ($\omega_{1,2}$) and
$Q$-factors of the circuits in comparison with the results shown
in Figure 3a. Additionally the degree of cross-talk is now greatly
increased as shown in the inset to Figure 3b. In particular, for
the case where SET1 is operated in the over-matched regime, a
poorly tuned stub network can create cross-talk signals as large
as 50\%.
\begin{figure}[t!]
\begin{center}
\includegraphics[width=8.5cm]{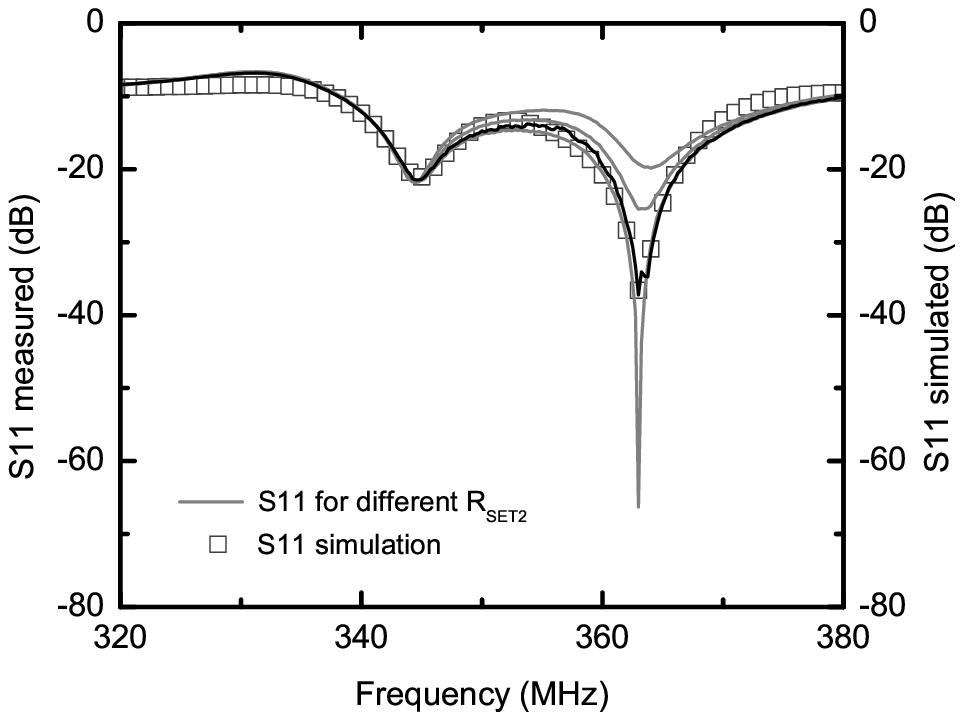}
\caption{Comparison between simulation results and
experimental data. For the experimental data $L_1$=780nH,   
$L_2$=660nH, $C$=0.3pF, $\omega_1$=345MHz, $\omega_2$=364MHz,
R$_{SET1}>$10M$\Omega$ and R$_{SET2}$ varies from $>$10M$\Omega$
(top) to $\sim$43k$\Omega$ (bottom). We adjust the lengths of the
stub network and the length of the bond wires linking the chip to
the circuit board as fitting parameters. Small deviations between
the simulation results and experiment are most likely due to the 
input impedance of the cryogenic amplifier, which deviates from  
50$\Omega$ across the frequency span.}
\vspace{-0.5cm}
\end{center}
\end{figure}

The above discussion, in which the length of transmission line
between the `T' section and the SET tank circuits strongly affects
both matching and cross-talk, is valid in the regime where line
lengths approach or exceed $\lambda/4$, which at these frequencies
is the order of centimeters. Although the use of micro-strip or
strip-line techniques can be used to avoid this issue, we note
that the constraint of independent source-drain biasing for each
SET requires the incorporation of a bias-tee in the transmission
line in between the `T' junction and SETs. Bias-tees typically
contain several inductors of appreciable size and are not easily
miniaturized with low insertion loss and good standing wave
ratios. Of further interest, a stub network (or double-stub) could
be used as an impedance matching technique without the requirement
of chip inductors commonly used for impedance matching in the
rf-SET. With the addition of a varactor diode terminating the
stub, the value of the reactance provided by the stub can be
varied with an applied voltage. Such a technique would have
several advantages over the present lumped-element matching
technique, including {\it in-situ} control of the matching and
$Q$-factor.

We now compare our simulation results with experimental data.
Figure 4 shows $S11$ data taken with a network analyzer for
varying SET2 resistances, $\sim$43k$\Omega$ (minimum S11 at
364MHz) to $>$10M$\Omega$ (maximum S11 at 364MHz). The resistance
of SET1 is kept constant throughout (R$_{SET1}\sim$10M$\Omega$,
biased in the superconducting gap). Using the microwave simulation
package we are able to produce $S11$ traces that approximate our
experimental results. As fitting parameters we adjust the length
of transmission lines $d_1$ and $d_2$ in the simulation, since
these are difficult to measure experimentally at mK temperatures.
The values used here ($d_{1}$=313mm, $d_{2}$=304mm) in the
simulation results however, are in agreement with what we expect
for our setup. In this arrangement the cross-talk between the tank
circuits is limited to $<$ 5\%, which is  close to its minimum
value for this frequency separation and the SET resistance range.
For the simulation results we also tune the length of the bond
wire between our circuit board and chip, and the input impedance
of our cryogenic microwave amplifier \cite{Berkshire}. The
amplifier impedance deviates from the ideal case of 50$\Omega$
since it contains an input network to achieve noise matching to
the first-stage transistor. Simultaneous noise and impedance
matching are generally not achieved, with the result being a
reduction in signal level at the output in order to maximize the
signal-to-noise ratio \cite{hagan}. Further weak deviations
between our experimental data and the simulation results are most
likely due to an inaccurate frequency dependence of the simulated
amplifier input impedance and the variation in parameters
characterizing our circuit elements with temperature.

\section{Sensitivity and charge noise rejection}

We now present measurement results for simultaneous operation of
two rf-SETs, beginning with the charge sensitivities. The data
shown in Figure 5 was taken with the SETs in the superconducting
regime. Operating the twin rf-SET in the superconducting mode
maximizes the device sensitivity, since the transconductance and
small signal differential conductance are maximized. Figure 5a
shows frequency domain data for both rf-SETs in response to a 2.5
MHz sine wave signal applied to a gate with a rms amplitude of
$\sim 0.1e$ on each SET island. Clear amplitude modulation (AM) of
both rf carriers is observed. The sensitivity of the rf-SET can
now be determined by measuring the signal to noise ratio (SNR) of
either sideband for a rms induced charge signal $q_{rms}$ on the
SET island. The spectral density of the noise is calculated as
follows \cite{Aassime_APL}:

\begin{equation}
\delta q = \frac{q_{rms}}{\sqrt{B} \times 10^{SNR/20}}
(e/\sqrt{Hz})
\end{equation}

where $B$ is the chosen resolution bandwidth of the spectrum
analyzer. For the twin rf-SET device studied here, we measure an
optimum sensitivity of $\delta q = 7.5 \mu e / \sqrt{Hz}$ and
$\delta q = 4.4 \mu e / \sqrt{Hz}$ for SET1 and SET2  with
resonances at 335MHz and 360MHz respectively. The sensitivities
were measured for $q = 0.005 e$ signals at 1.3 MHz in the dc+rf
(superconducting) mode \cite{Aassime_APL}. The frequency
dependence of the noise floor shown in Figure 5a is primarily
linked to the frequency dependence of the noise temperature of the
cryogenic amplifier, which reaches a minimum of $T_N \sim 1.5$K at
$\omega \sim$350MHz. Additional (weak)  contributions include shot
noise associated with the SET current, phase noise of our rf
sources, $1/f$ noise in the SETs  and electronics and the
frequency dependent noise of the room temperature amplifiers and
mixers.

\begin{figure}[t!]
\begin{center}
\includegraphics[width=8.5cm]{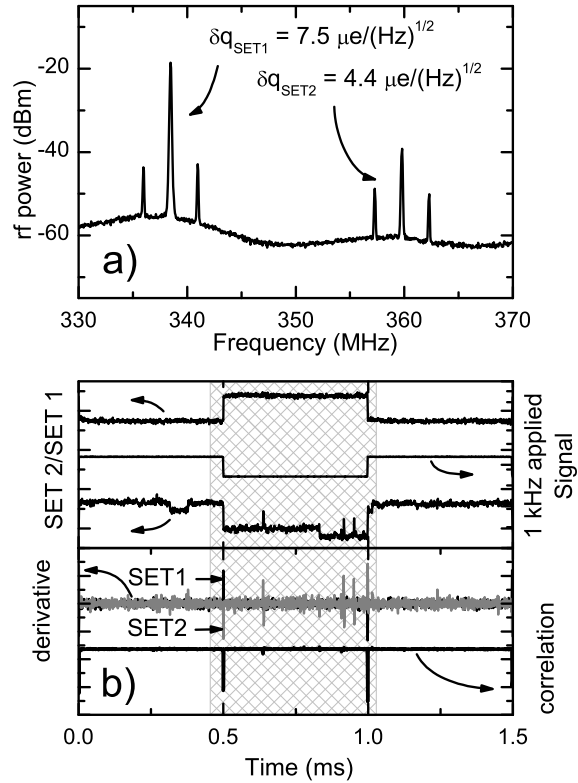}
\caption{\textbf{a)} Typical amplitude modulation (AM)
signal associated with both the SETs responding to a $\sim 0.1e$,
2.5MHz gate signal. Charge sensitivities for each device are
stated for the case of simultaneous operation (different
measurement). \textbf{b)}  time domain data taken simultaneously
with both SETs. Both devices respond independently to a 1KHz
square wave signal applied to a nearby gate. Spurious charge noise
is superimposed on the output of SET2 (shown in the hashed
region). By cross-correlating the signals from both SETs  
(multiplying the derivatives) we are able to suppress spurious
charge noise events in real time.}
\vspace{-0.5cm}
\end{center}
\end{figure}

Focusing on the response of the SETs in the time domain, we now
demonstrate the suppression of charge noise via cross-correlated
measurements. Figure 5b shows the response of both SETs (measured
simultaneously) when a 1 kHz square wave is applied to a nearby
gate via a 40 dB attenuator. This signal induces charges of $\sim
0.1 e$ and $\sim 0.05 e$ on the islands of SET1 and SET2
respectively. Although the output of the two SETs faithfully
follow the gate signal, additional unwanted charge noise is also
picked up by SET2 (e.g. in the 0.7 - 1 ms time interval). The
observed random telegraph signals (RTSs) are probably associated
with charge noise arising from two-level charge traps in the SET
oxide tunnel barrier or substrate. Using a multi-channel digital
oscilloscope, we are able to multiply the time derivative signals
from SET1 and SET2 in {\it real-time}. The result is shown in the
lower portion of Figure 5b and constitutes a real-time
cross-correlation of the SET signals. Close inspection of the
correlation trace shows that this technique produces sharp spikes
when there is a true signal that affects both SETs (such as the
rising or falling edges of the square wave applied to the gate)
and a clear suppression of the random charge noise by up to 95\%
of the peak height correlation. This ability to reject events
associated with charge noise is likely to  be important for the
application of SETs as detectors and in the readout of solid state
quantum computers.

\section{Conclusion}

In conclusion we have described the development of a twin rf-SET
device that makes use of two tuned impedance transformers to
perform wavelength division multiplexing (WDM) using a single
cryogenic following amplifier. In conjunction with a mixer-based
demodulation technique, we have explored the issues affecting
cross-talk and sensitivity. The twin rf-SET has been shown to
perform real time cross-correlation of SET signals. Such a
correlation technique suppresses charge noise and enables true
signals associated with readout to be distinguished from spurious
artifacts on time-scales required for solid state quantum
computation. Future work will investigate the use of the twin
rf-SET for charge motion detection on microsecond time-scales, and
its application to readout of both semiconducting and
superconducting charge qubits.

We thank D. Barber for technical support and S. Kenyon, K. Lehnert
and R. Schoelkopf for fruitful discussions and insights. This work
was supported by the Australian Research Council, the Australian
Government and by the US National Security Agency (NSA), Advanced
Research and Development Activity (ARDA) and the Army Research
Office (ARO) under contract number DAAD19-01-1-0653. DJR
acknowledges a Hewlett-Packard Fellowship.


\end{document}